\documentclass{ws-rv9x6}
\usepackage{ws-rv-van}     % numbered citation/references (default)
\begin{document}

\chapter[Pairing fluctuations in rotating nuclei]{
Pairing fluctuations and gauge symmetry restoration
in rotating superfluid nuclei
}

\author[Y.~R.~Shimizu]{
Yoshifumi R. Shimizu
}
%\index[aindx]{Author, F.} % or \aindx{Author, F.}
%\index[aindx]{Author, S.} % or \aindx{Author, S.}

\address{
Department of Physics, Faculty of Sciences, Kyushu University, \\
 Fukuoka 812-8581, Japan
}

\begin{abstract}
Rapidly rotating nuclei provide us good testing grounds
to study the pairing correlations;
in fact, the transition from the superfluid to the normal phase
is realized at high-spin states.
The role played by the pairing correlations is quite different
in these two phases:
The static (BCS like mean-field) contribution is dominant
in the superfluid phase,
while the dynamic fluctuations beyond the mean-field approximation
are important in the normal phase.
The influence of the pairing fluctuations
on the high-spin rotational spectra and moments of inertia is discussed.
\end{abstract}

\body

\section{Rotation and pairing correlations $-$ Introduction}
\label{sec:intro}

In all the different contributions in this Volume,
various aspects of the pairing correlations,
which play important roles in nuclear physics, are discussed.
In this contribution I would like to concentrate on
the effect of the pairing fluctuations
in rapidly rotating nuclei\cite{SGB89},
which is generic and yet far from trivial.
Here the pairing fluctuations mean the dynamic motions
of the pairing gap, i.e., so-called the pairing vibrations\cite{BB66}
(see also Ref.~\citen{BHR73}), whose effects appear beyond
the static (BCS) mean-field approximation and are characteristic
in the finite system like atomic nucleus.

It is well known that most of non-closed shell nuclei,
which have quadrupole deformed shape,
exhibit collective nuclear rotations\cite{BMtext}.
In the 80's,
the combined developments of the heavy-ion accelerators
and the high-resolution $\gamma$-ray detectors made it possible
to explore the properties of rapidly rotating nuclei,
i.e., the high-spin states up to spin values $I \approx 60\hbar$,
of medium and heavy nuclei.
Many interesting phenomena and issues
have been revealed; see e.g. Refs.~\citen{VDS83,Szym84,GHH86},
and Refs.~\citen{Fra01,SW05} for more recent progress.
The pairing correlations, either static or dynamic, play a crucial role
in most of these phenomena over the wide spin range.
As for a well-known example, the superfluidity
is responsible for the reduction of the moment of inertia
for the collective nuclear rotation near the ground state\cite{BMP58};
it takes only about (or even less than) half of the rigid-body value,
which is expected for
the independent nucleonic motions in the deformed mean-field.
In this way, the ground states of deformed nuclei can be well described
by the BCS theory with finite pairing gaps, ${\mit\Delta}\approx 1$ MeV,
and the BCS quasiparticles appear as a basic excitation mode.
In fact, the ``backbending'' phenomenon\cite{JRS71},
which is systematically observed at spin $I\approx 10-16\hbar$
in the yrast\footnote{
 The word ``yrast'' means dizziest, and the yrast state is
 the lowest energy state at a given angular momentum.
 Connecting the yrast states composes the yrast band.}
bands of medium and heavy nuclei, can be understood as a band-crossing
between the BCS vacuum and a specific two-neutron-quasiparticle
excited configuration that is particularly favored by
the effect of rotation\cite{SS72}
(see the contribution of F.~Stephens and I.-Y.~Lee and that of P.~Ring
to this volume).
After the understanding of this novel phenomenon,
it was realized that not only the yrast band but also
many excited rotational bands can be well described by
the concept of independent quasiparticle excitations
in the rotating frame\cite{BF79}.  This is quite nontrivial;
complex rotational spectra at high-spin states can be nicely described
by the so-called cranked shell model~\cite{BFM86}
(see the contribution of S.~Frauendorf to this Volume),
which is one of the most important achievements
in the studies of rapidly rotating nuclei.

The Cooper pair in nucleus is composed of a pair of nucleons
in the time-reversal conjugate orbits whose angular momenta
couple to $J=0$.\footnote{
 The nucleon pairs with higher multipole,
 e.g. the quadrupole pair ($J=2$),
 also play important roles especially in deformed nuclei.}
The effect of rotation, which appears
as the Coriolis and centrifugal forces in the rotating frame,
tends to align the angular momenta of nucleonic orbits
to the rotation axis, and consequently breaks the Cooper pairs.
In analogy to the metallic superconductors in the magnetic field,
it was predicted that the phase transition from the superfluid
to the normal phase is induced by the rapid rotation\cite{MV60}.
However, a sharp transition as in macroscopic systems would not be
expected in a finite system such as the nucleus.
Instead, the finite nuclear system provides the opportunities
to study a ``phase transition'' in terms of
the individual quantum states such as the rotational-band spectra
with non-negligible effects of the dynamic fluctuations.
In fact, the transition is not very simple even within
the mean-field approximation:
The effect of the band-crossings (backbendings),
i.e., the successive excitations (alignments) of quasiparticles,
is more dramatic\cite{SM83},
and the calculated pairing gap reduces stepwisely
along the yrast states.  It is now believed that
the unpaired phase is realized for neutrons
at spins $I\approx 20-30\hbar$ in the rare-earth region,
evidence for which is given by comparing the observed spectra
with the rotating single-particle energies
with zero pairing gap\cite{Garr85,BEG85}.
However, it was recognized that the effects of pairing correlations
remain considerably after vanishing
the static (BCS) pairing gap\cite{ERI85,BDF86};
the ``effective pairing gap'' including the dynamic fluctuations
beyond the static mean-field does not vanish and
only gradually decrease across the phase transition\cite{SGB89,ERI85}.

In the following, after briefly reviewing how to treat
the nuclear rotational motion,
I discuss the theoretical method to evaluate the pairing fluctuations
within the random phase approximation (RPA)\cite{FWtext,RStext}.
A few examples of the calculated results,
taken from our studies in Refs.~\citen{SGB89,SB88,SVB90,SB90,YRS90,SDB00},
are presented in comparison with experimental data.

\section{Description of rotational motion $-$ Cranking model}
\label{sec:descrot}

In order to make this article self-contained,
here I recapitulate the method to treat the rotational motion
and to analyse the rotational spectra;
see e.g. Refs.~\citen{VDS83,Szym84,BF79,BFM86}
for detailed accounts.

The nuclear collective motion is treated semiclassically,
which is called the ``cranking'' prescription\cite{Ing54}.
Namely the Hamiltonian of the system is transformed
into the {\it uniformly} rotating frame\footnote{
 $\hbar=1$ unit is used for mathematical expressions.},
\begin{equation}
 \hat H' = \hat H - \omega_{\rm rot} \hat J_x,
\label{eq:crank}
\end{equation}
where $\omega_{\rm rot}$ denotes the rotational frequency
about the rotation axis ($x$-axis), which is chosen to be one of
the principal axes of the deformed body with largest moment of inertia,
and is usually perpendicular to the symmetry axis
of the quadrupole nuclear shape.
Since we are mainly interested in the lowest energy (yrast) high-spin states,
this is a natural assumption
(see Ref.~\citen{Fra01} for more general situations).
The energy in the rotating frame $E'=\langle H'\rangle$,
which is called the ``routhian'',
and the angular momentum along the rotation axis,
$I_x=\langle \hat J_x \rangle=-{\partial E'}/{\partial \omega_{\rm rot}}$
with $I_x=\sqrt{I(I+1)}\approx I+\frac{1}{2}$,
are evaluated as functions of
the rotational frequency $\omega_{\rm rot}$.

On the other hand, the nuclear collective rotation is measured as
the rotational spectra, $E(I)$, which are composed of
a group of states with different angular momentum $I$ changing by two units
(${\mit\Delta}I=2$), and connected by the strong
electric quadrupole ($E2$) $\gamma$-ray emissions.
In accordance with the simple assumption of rotational motion
in Eq.~(\ref{eq:crank}), the rotational frequency is calculated by
\begin{equation}
 \omega_{\rm rot}(I)=\frac{\partial E}{\partial I}
 \approx \frac{E(I+1)-E(I-1)}{(I+1)-(I-1)}=\frac{1}{2}E_\gamma,
\label{eq:wrot}
\end{equation}
with the $\gamma$-ray energy $E_\gamma$
of the associated rotational transition.
This implicitly defines the relation $I_x(\omega_{\rm rot})$,
between the angular momentum $I_x$
and the rotational frequency $\omega_{\rm rot}$,
and then the experimental routhian $E'(\omega_{\rm rot})$ is obtained as
\begin{equation}
 E'(\omega_{\rm rot}) = E(I(\omega_{\rm rot}))
 - \omega_{\rm rot}I_x(\omega_{\rm rot}).
\label{eq:routh}
\end{equation}
In this way the theoretical routhians can be directly
compared with the experimental routhians,
although the latter are given only at the discrete points
of the rotational frequencies.

In the mean-field approximation, e.g., in the cranked shell model,
the Hamiltonian $\hat H$ is replaced with the one-body Hamiltonian,
\begin{equation}
 \hat H \quad\rightarrow\quad
 \hat h = \hat h_{\rm def} - {\mit\Delta}(\hat P^\dagger+\hat P)
  -\lambda \hat N,
\label{eq:mf}
\end{equation}
where $\hat h_{\rm def}$ describes the single-particle motion
in the deformed average potential, the second term is
the pair-field with $\hat P^\dagger$ being
the monopole pair creation operator,
\begin{equation}
 \hat P^\dagger = \frac{1}{2} \sum_{i}
 \hat c^\dagger_i \hat c^\dagger_{\tilde i}
 \quad (\mbox{$\tilde i$: time reversed orbit of $i$}),
\label{eq:monoP}
\end{equation}
and the last term $-\lambda \hat N$ ensures the correct particle number
on average, because the number conservation is broken in the BCS treatment.
By diagonalizing the cranking Hamiltonian with Eq.~(\ref{eq:mf})
the quasiparticle energies in the rotating frame are obtained,
which can be directly compared with the complex rotational spectra
for both even and odd nuclei\cite{BFM86};
see the contribution of S.~Frauendorf to this volume for detailed explanations.
Of course, it can be well used with ${\mit\Delta}=0$ for
the case of quenched pairing correlations,
i.e., for the normal phase routhians.

\section{Pairing fluctuations with RPA method}
\label{sec:pfluc}

The dynamic pairing fluctuations beyond the mean-field approximation
is induced by the two-body interaction.
The simple one, the so-called monopole pairing force, is employed
with the operator $\hat P^\dagger$ defined in Eq.~(\ref{eq:monoP});
\begin{equation}
 \hat H = \hat h_{\rm def}+ \hat V,\qquad
 \hat V=-\frac{G}{2}
 \bigl(\hat P^\dagger \hat P + \hat P \hat P^\dagger\bigr),
\label{eq:mpforce}
\end{equation}
with the strength $G$.
The BCS treatment of this Hamiltonian leads to the one-body Hamiltonian
in Eq.~(\ref{eq:mf}) with the {\it selfconsistent} (static) pairing gap
${\mit\Delta}=G\langle \hat P^\dagger \rangle_{\rm mf}
=G\langle \hat P \rangle_{\rm mf}$, which is nothing else but
the order parameter of the super-to-normal phase transition.

The fluctuations about the mean-field are calculated by
diagonalizing the Hamiltonian~(\ref{eq:mpforce}) within the RPA.
The induced energy gain is given by
\begin{equation}
 E_{\rm corr}^{\rm RPA}=\frac{1}{2}\biggl[
  \sum_n \omega_n - \sum_{\alpha>\beta}(e_\alpha+e_\beta)\biggr],
\label{eq:Ecorr}
\end{equation}
where $\omega_n$ is the RPA eigenenergy and $e_\alpha$ is
the quasiparticle (particle or hole) energy in the superfluid (normal) phase.
They are calculated with the cranking prescription~(\ref{eq:crank})
as functions of $\omega_{\rm rot}$ to study the rapidly rotating nuclei.
Thus, the total RPA routhian is calculated as
\begin{equation}
 E'_{\rm RPA}=E'_{\rm mf}+E_{\rm corr}^{\rm RPA},
 \quad
 E'_{\rm mf}=\langle \hat h_{\rm def}-\omega_{\rm rot}\hat J_x\rangle_{\rm mf}
 -G\langle \hat P^\dagger\rangle^2_{\rm mf}.
\label{eq:RPArouth}
\end{equation}
It should be mentioned that $E_{\rm corr}^{\rm RPA}$ in Eq.~(\ref{eq:Ecorr})
contains the exchange energy,
$E_{\rm ex}=\langle \hat V \rangle_{\rm mf}
+G\langle \hat P^\dagger\rangle^2_{\rm mf}$,
which is found to be rather constant\cite{SGB89} against
the change of $\omega_{\rm rot}$.
Note that the calculation of $E_{\rm corr}^{\rm RPA}$ requires
all the RPA eigenenergies, which amount to a few or more than ten
thousands depending on the pairing model space.
Since the convergence with respect to the number of solutions
is slow\cite{EMR80},
it is important to include all of them for stable results,
which is a numerically demanding task.
A general efficient method to perform the calculation
was developed in Ref.~\citen{SGB89}
by utilizing the linear response theory,
and it was further improved in Ref.~\citen{SDB00}.

\subsection{Response function technique}
\label{subsec:resp}

Generally the two-body interaction can be represented
by the form of multi-component separable force,
\begin{equation}
 \hat V=-\frac{1}{2}\sum_{\rho=1}^q \chi_\rho \hat Q_\rho \hat Q_\rho,
 \qquad Q_\rho^\dagger=Q_\rho,
\label{eq:separable}
\end{equation}
with Hermitian one-body operators $\hat Q_\rho$ and
strengths $\chi_\rho$  $(\rho=1,2,...,q)$.
For the monopole pairing interaction~(\ref{eq:mpforce}),
$q=2$ and
$\hat Q_1 \equiv\hat P^\dagger+\hat P$,
$iQ_2 \equiv\hat P^\dagger-\hat P$,
and $\chi_1=\chi_2 \equiv G/2$.
The RPA eigenvalue problem can then be replaced to solve
the following dispersion equation,
\begin{equation}
 \det{\cal R}(\omega)=0,
 \quad\mbox{with}\quad
 {\cal R}(\omega)=[1-R(\omega)\chi]^{-1}R(\omega),
\label{eq:disp}
\end{equation}
where the $q \times q$ matrices, ${\cal R}(\omega)$ and $R(\omega)$,
are composed of the RPA and the unperturbed response functions
for the operators $\hat Q_\rho$,
and the diagonal matrix $\chi=(\delta_{\rho\sigma}\chi_\rho)$;
\begin{equation}
 R_{\rho\sigma}(\omega)
 \equiv \sum_{\alpha>\beta}\biggl[
  \frac{q^*_\rho(\alpha\beta)q_\sigma(\alpha\beta)}{e_\alpha+e_\beta-\omega}
 +\frac{q_\rho(\alpha\beta)q^*_\sigma(\alpha\beta)}{e_\alpha+e_\beta+\omega}
 \biggr],
\label{eq:respfn}
\end{equation}
with $q_\rho(\alpha\beta)
\equiv \langle \alpha\beta|\hat Q_\rho|0\rangle_{\rm mf}$.
Then, by employing the adiabatic turn-on the interaction
and the analytic property of the response function~(\ref{eq:respfn}),
it was shown that the correlation energy
can be calculated by the following formula\cite{SDB00},
\begin{equation}
 E_{\rm corr}^{\rm RPA} = \frac{1}{2\pi}\int_0^\infty
 {\rm Re}\bigl[ \log\{\det[1-R(i\omega)\chi]\} \bigr]\,d\omega,
\label{eq:respEcorr}
\end{equation}
so that it is {\it not} necessary to explicitly solve Eq.~(\ref{eq:disp}).
Note that the integration is taken
along the upper imaginary axis $z=i\omega$ in the complex energy plane,
for which the integrand is a smoothly decreasing function
and the numerical integration can be done straightforwardly.
In Ref.~\citen{SGB89} a different integration path is taken near the positive
real axis, where the integrand is a oscillating function,
and the numerical integration should have been done more carefully
(see Ref.~\citen{SGB89,SDB00} for detailed discussions).

It is instructive to consider
the following RPA pairing gap~\cite{SGB89,SB90,SDB00},
\begin{equation}
 {\mit\Delta}_{\rm RPA} = G\sqrt{\frac{1}{2} \sum_n
 \Bigl[\langle 0 | \hat P^\dagger | n \rangle
 \langle n | \hat P | 0 \rangle + 
 \langle 0 | \hat P | n \rangle
 \langle n | \hat P^\dagger | 0 \rangle\Bigr]_{\rm RPA} },
\label{eq:RPAgap}
\end{equation}
in keeping with ${\mit\Delta}_{\rm NP}$ introduced in the variation
after number projection (NP) approach\cite{ERI85}
(see \S\ref{subsec:NP}).
However, the contribution of the zero mode, i.e., the symmetry recovering
Nambu-Goldstone (NG) mode (the pairing rotation\cite{BHR73}),
which is present in the superfluid phase,
diverges because of the small amplitude approximation inherent in the RPA:
It is natural to replace its (divergent) contribution to that of the mean-field;
\begin{equation}
 \frac{1}{2}\Bigl[|\langle 0 | \hat P^\dagger | n \rangle|^2
 + |\langle 0 | \hat P | n \rangle\bigr|^2\Bigl]_{n={\rm NG}}
 \ \rightarrow\quad
 \langle \hat P^\dagger \rangle_{\rm mf}\langle \hat P \rangle_{\rm mf}
 =({\mit\Delta}/G)^2.
\label{eq:NGgap}
\end{equation}
Then the difference between the squared RPA and mean-field pairing gaps,
${\mit\Delta}^2_{\rm RPA}-{\mit\Delta}^2$, represents the effect of
pairing vibrations, which can be calculated by integrating the trace
of the RPA response matrix $\mbox{Tr}{\cal R}(\omega)$
without explicitly solving the RPA equation\cite{SGB89,SB90,SDB00}.

\begin{figure}[htb]
\centerline{\epsfig{file=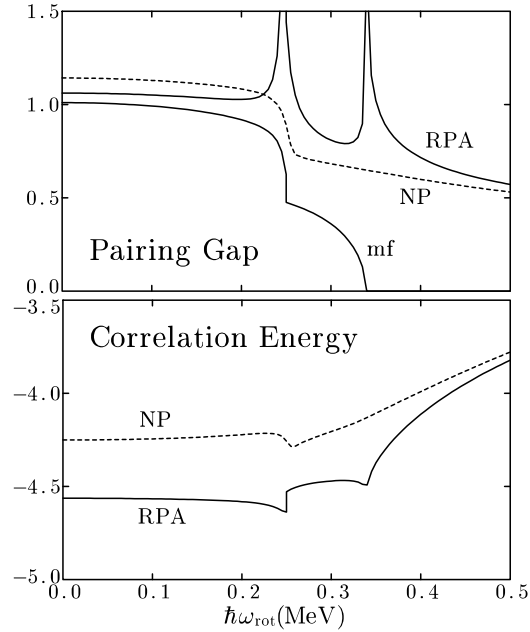,width=7cm}}
\vspace*{-2mm}
\caption{The RPA and BCS mean-field (mf) pairing gaps (upper)
and the RPA correlation energies (lower) for neutrons
in the yrast band of $^{164}$Er
as functions of the rotational frequency $\omega_{\rm rot}$.
The results by the variation after number projection (NP) method
are included as dashed lines.
Here the exchange contributions are excluded\cite{ERI85,SB90}
both in ${\mit\Delta}_{\rm RPA}$ and ${\mit\Delta}_{\rm NP}$.
Taken from Ref.~\citen{SDB00} with eliminating two irrelevant lines.
}
\label{fig1}
\vspace*{-2mm}
\end{figure}

In Fig.~\ref{fig1}, an example of the RPA correlation
energy and the pairing gaps are shown.
The mean-field pairing gap ${\mit\Delta}$ reduces stepwisely to zero
at the critical frequency $\omega_{\rm rot}=\omega_{\rm c}\approx 0.33$ MeV
of the super-to-normal phase transition.
The first reduction at $\omega_{\rm rot}\approx 0.24$ MeV is caused
by the two-neutron-quasiparticle alignments (excitations)
related to the backbending phenomenon,
where the correlation energy $E_{\rm corr}^{\rm RPA}$
is discontinuous.  At $\omega_{\rm rot}=\omega_{\rm c}$ it is continuous
but its derivative, i.e., the correction to the alignment,
$\delta I_x=-{\partial E_{\rm corr}^{\rm RPA}}/{\partial \omega_{\rm rot}}$,
diverges, which is a drawback of the RPA and one has to
go beyond the RPA\cite{SB88}
or to make smooth interpolations to compare with the experimental data.
At these two frequencies the RPA gap ${\mit\Delta}_{\rm RPA}$
diverges, because one of the RPA eigenenergies goes across zero.
It should be mentioned that $E_{\rm corr}^{\rm RPA}$
is almost constant as long as the BCS pairing gap is sizable,
while its absolute value decreases after its quenching;
therefore the effect of $E_{\rm corr}^{\rm RPA}$ is important
after the static pairing gap becomes small.
In contrast to the mean-field gap, ${\mit\Delta}_{\rm RPA}$
keeps finite values even at highest frequencies,
reflecting that the pairing fluctuations remains considerably
in the normal phase.
These behaviors of the correlation energy
and the pairing gaps are rather general in rapidly rotating nuclei\cite{SGB89}.

\subsection{Routhians and alignments in normal deformed nuclei}
\label{subsec:ND}

\vspace*{-4mm}

\begin{figure}[htb]
\centerline{\epsfig{file=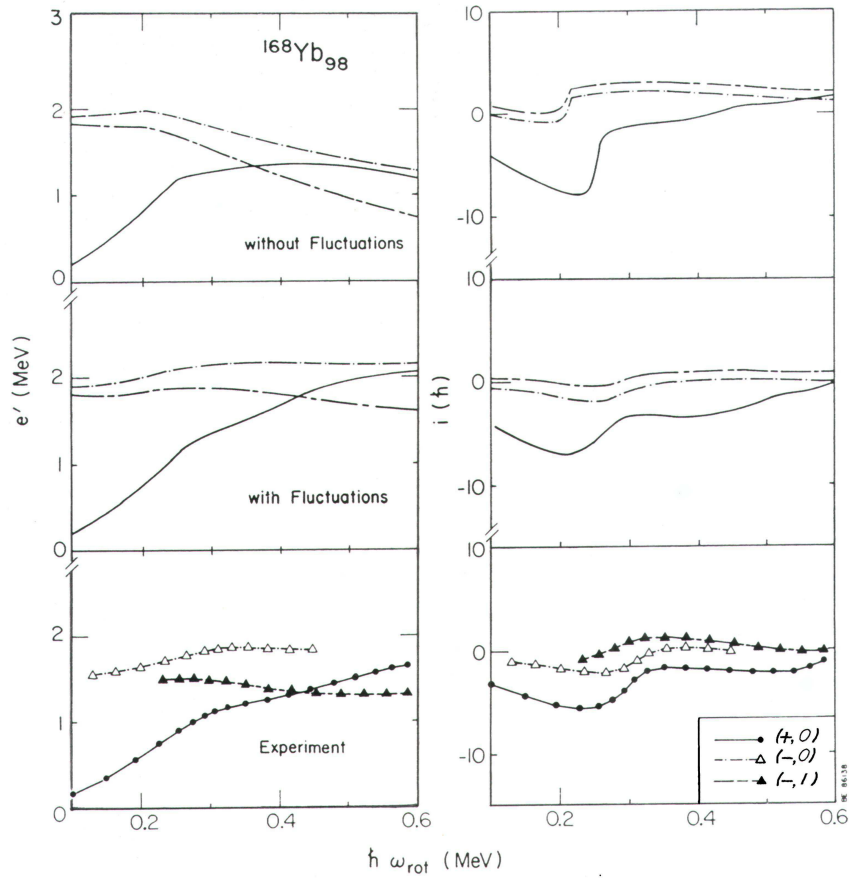,width=11.5cm}}
\vspace*{-2mm}
\caption{Calculated and experimental routhians $e'$ (left-hand side)
and alignments $i$ (right-hand side)
for the lowest three configurations in $^{168}$Yb;
the top panels display the calculation without fluctuations,
middle with fluctuations, and bottom experimental data.
Taken from Ref.~\citen{SGB89}.}
\label{fig2}
\vspace*{-2mm}
\end{figure}

In order to discuss how the correlation energy affects rotational spectra 
at high-spin states, the routhians $e'$
and the aligned angular momenta (``alignments'') $i$
for the lowest three configurations in the nucleus $^{168}$Yb
are shown in comparison with experimental data in Fig.~\ref{fig2}.
Here these quantities are plotted relative to the so-called
rigid-body reference, i.e.,
\begin{equation}
 e'(\omega_{\rm rot}) = E'(\omega_{\rm rot})
 + \frac{1}{2}{\cal J}_0\, \omega_{\rm rot}^2,
 \quad
 i(\omega_{\rm rot}) = I_x(\omega_{\rm rot})
 - {\cal J}_0\, \omega_{\rm rot},
\label{eq:relei}
\end{equation}
where ${\cal J}_0$ is the rigid-body moment of inertia.
In this calculation with a rather simple interaction in Eq.~(\ref{eq:mpforce})
the experimental moment of inertia cannot be described correctly,
and the ${\cal J}_0$ value for theoretical results is adjusted
so as to reproduce the lower frequency part of routhians $e'$,
for which the correlation energy remains almost constant
as it is shown in Fig.~\ref{fig1}.
Apparently, the calculated routhians of the higher frequency part is
smaller than the experimental data without the pairing fluctuations.
The $(+,0)$ configuration is the band with positive parity
and even spins and corresponds to the yrast band;
the kink of its routhian at $\omega_{\rm rot}\approx 0.28$ MeV
corresponds to the two-neutron-quasiparticle crossing.
In this band the mean-field pairing gap almost quenches around
$\omega_{\rm rot}\approx 0.4$ MeV, and certainly the effect
of the pairing fluctuations becomes more evident at larger frequencies.
The negative parity excited bands with $(-,0)$ and $(-,1)$
are two neutron excited configurations
and their pairing gaps are about 60\% of the $(+,0)$ band at lowest frequency.
Therefore, their static pairing correlations are reduced more
than that of the $(+,0)$ band, and the effects of pairing fluctuations
are more conspicuous in the relatively lower frequency region.
With these effects of the fluctuations,
the overall agreement between the calculation and
the experiment apparently improves.
From the general dependence of $E_{\rm corr}^{\rm RPA}$ on $\omega_{\rm rot}$,
the correction to the alignments $i$ is always negative,
which is called ``dealignment'', and it amounts to $2-3\hbar$;
again, this makes the agreement of alignments much better.

\subsection{Moments of inertia in superdeformed nuclei}
\label{subsec:SD}

It was also discussed\cite{SVB90} that
the pairing fluctuations play important roles
in the nuclei with very large deformation,
which are called ``superdeformation'' and very regular rotational bands
have been systematically observed; see Refs.~\citen{NT88,JK91}
and the contribution of P.-H.~Heenen to this Volume.

\begin{figure}[htb]
\centerline{\epsfig{file=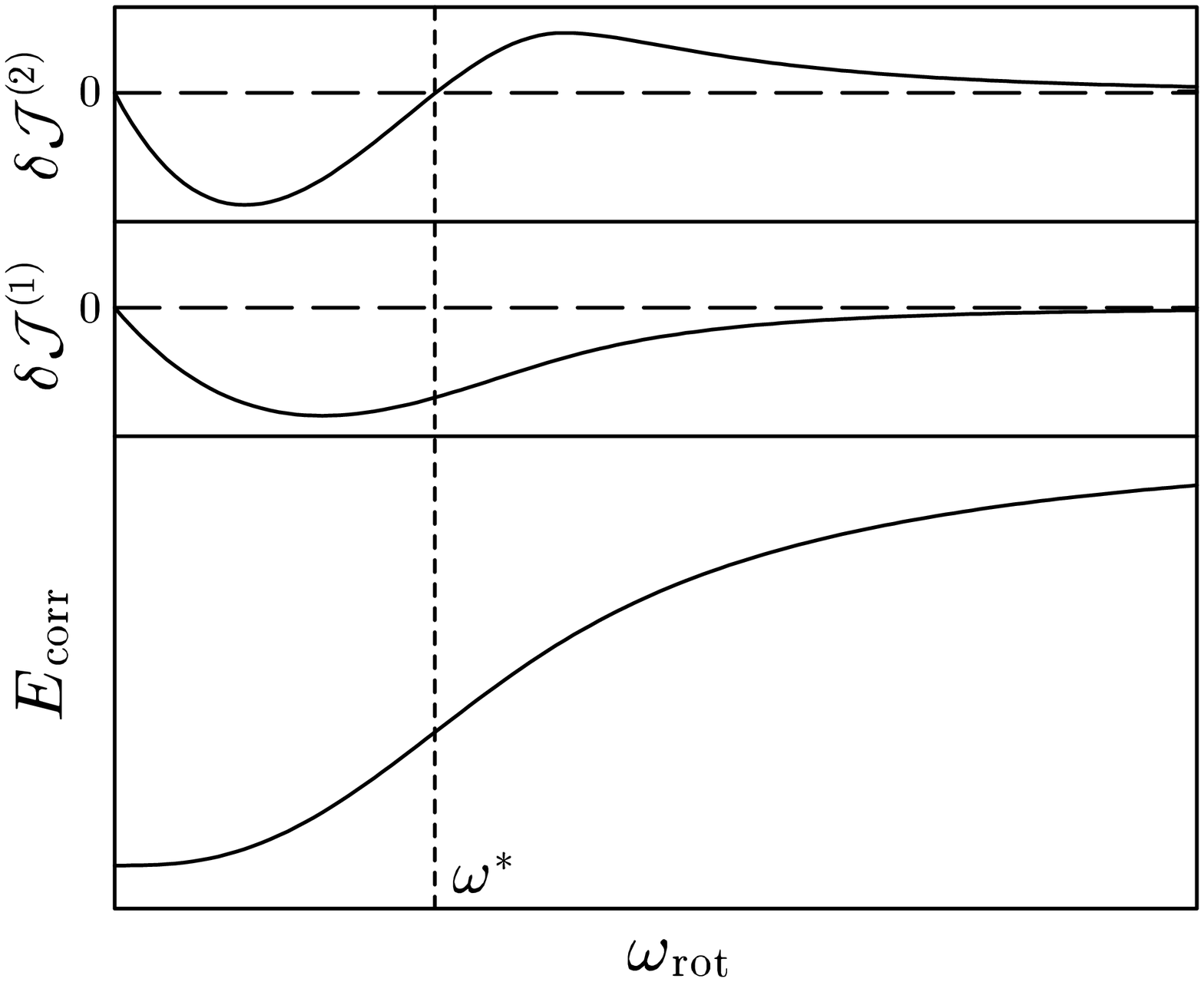,width=6cm}}
\vspace*{-2mm}
\caption{Schematic figure depicting the (smoothed)
pairing correlation energy and its influence
on the two moments of inertia ${\cal J}^{(1)}$ and ${\cal J}^{(2)}$.}
\label{fig3}
\vspace*{-2mm}
\end{figure}

\begin{figure}[htb]
\centerline{\epsfig{file=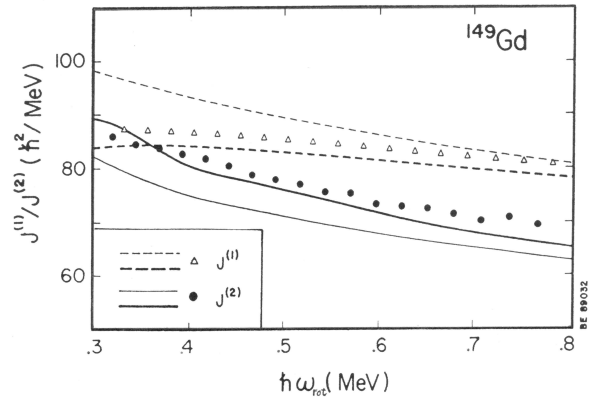,width=8cm}}
\vspace*{-2mm}
\caption{Calculated and experimental moments of inertia
${\cal J}^{(1)}$ and ${\cal J}^{(2)}$ for the yrast superdeformed band
in $^{159}$Gd; thick (thin) lines denote the result
with (without) pairing fluctuations by using the RPA method.
Taken from Ref.~\citen{SVB90}.}
\label{fig4}
\vspace*{-2mm}
\end{figure}

For the analysis of these superdeformed bands,
the two moments of inertia are utilized quite often;
they are called the kinematic and dynamic inertia,
${\cal J}^{(1)}$ and ${\cal J}^{(2)}$, respectively, and are defined by
\begin{equation}
{\cal J}^{(1)}\equiv
\frac{I_x}{\omega_{\rm rot}}=
-\frac{1}{\omega_{\rm rot}}\frac{\partial E'}{\partial\omega_{\rm rot}},
\quad
{\cal J}^{(2)}\equiv
\frac{\partial I_x}{\partial \omega_{\rm rot}}=
-\frac{\partial^2E'}{\partial^2\omega_{\rm rot}}.
\label{eq:J12}
\end{equation}
The corrections induced by the pairing fluctuations to these inertia,
$\delta {\cal J}^{(1)}=
-(1/\omega_{\rm rot})(\partial E_{\rm corr}/\partial\omega_{\rm rot})$
and
$\delta {\cal J}^{(1)}=
-\partial^2 E_{\rm corr}/\partial^2\omega_{\rm rot}$,
are schematically depicted in Fig.~\ref{fig3}.
Here the symbol $\omega^*$ denotes the frequency of the inflection point
in the correlation energy and is located, in most cases,
near the critical frequency of the pairing phase transition.

The superdeformed nuclei are realized by the strong shell effects
based on the special deformations, e.g., the integer axis ratio like $2:1$,
and reflect the characteristics of the deformed closed shell.
Therefore, just as in the case of the magic nuclei,
the pairing correlations are very much reduced.
Especially, those in the mass number $A \approx 150$ region
are believed to be in the normal phase (${\mit\Delta}=0$)
already in their lowest states,
and then the inflection point $\omega^*$ in Fig.~\ref{fig3} is expected
to be lower than the experimentally observed frequency region.
Thus, the corrections to the inertia is negative for ${\cal J}^{(1)}$
and positive for ${\cal J}^{(2)}$.  In Fig.~\ref{fig4},
the calculated and experimental inertia are compared
for the yrast superdeformed band of $^{149}$Gd.
The mean-field calculation overestimates ${\cal J}^{(1)}$
while it underestimates ${\cal J}^{(2)}$,
and a good agreement is obtained by including
the effects of pairing fluctuations;
again this is rather generic for superdeformed nuclei
in the $A\approx 150$ region\cite{SVB90}.

%with the variation after number projection method}
\subsection{Gauge symmetry restoration}
\label{subsec:NP}

In the RPA the broken symmetry is signalled by the appearance
of the zero-energy NG mode,
whose contribution to the correlation energy is largest,
$\approx 2{\mit\Delta}$, from Eq.~(\ref{eq:Ecorr}).
This implies the importance of restoring the gauge symmetry
(the number conservation);
the method is called the number projection, which explicitly
projects out wave functions with good particle numbers
from a gauge-symmetry broken wave function.
In fact, it has been known\cite{RStext} that the correlations
beyond the mean-field approximation can be taken into account
by optimizing the superfluid mean-field wave function
from which the projection is carried out;
the so-called variation after projection (VAP) approach.
Therefore the variation after number projection (NP)
is an alternative method to evaluate
the pairing fluctuations in the rotating nuclei
(see the contribution of J.~L.~Egido to this Volume).

\begin{figure}[htb]
\centerline{\epsfig{file=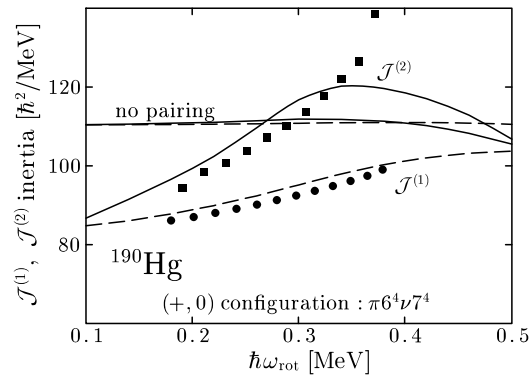,width=7cm}}
\vspace*{-2mm}
\caption{Calculated and experimental moments of inertia
${\cal J}^{(1)}$ and ${\cal J}^{(2)}$ for the yrast superdeformed band
in $^{190}$Hg by using the variation after number projection method.
Taken from Ref.~\citen{YRS90} but with newer experimental data.}
\label{fig5}
\vspace*{-2mm}
\end{figure}

Since the expectation value of the monopole pairing operator $\hat P^\dagger$
vanishes for the number conserving wave function,
the NP pairing gap is defined by the following\cite{ERI85};
\begin{equation}
 {\mit\Delta}_{\rm NP} = G\sqrt{\frac{1}{2}
 \bigl(\langle \hat P^\dagger \hat P \rangle_{\rm NP} + 
 \langle \hat P \hat P^\dagger \rangle_{\rm NP}\bigr) }.
\label{eq:NPgap}
\end{equation}
The correlation energy and the pairing gap evaluated by the NP approach
are also included in Fig.~\ref{fig1}.  It can be seen that both
quantities behave quite similarly to those evaluated by the RPA method,
although the NP correlation energy is smaller indicating
that the RPA method takes more correlations into account.
In Ref.~\citen{SB90} comparison of the RPA and NP methods were performed
for the routhians and alignments in rapidly rotating nuclei,
and it was found that indeed two methods give very similar results.
A merit of the NP method is that its result is smooth
across the critical point of the super-to-normal phase transition
in contrast to the RPA.  It is especially useful to calculate
the ${\cal J}^{(1)}$ and ${\cal J}^{(2)}$ inertias,
which require the first and second derivatives of the correlation energy.
An example is shown in Fig.~\ref{fig5},
where the NP method is applied\cite{YRS90}
to the yrast superdeformed band in $^{190}$Hg.
The ${\cal J}^{(2)}$ inertias of the superdeformed nuclei
in the mass number $A \approx 190$ region systematically show
increasing behaviors as $\omega_{\rm rot}$.
Because of smaller shell gaps than those in the $A\approx 150$ regions,
the stronger pairing correlations are expected.
Namely the inflection point $\omega^*$ in Fig.~\ref{fig3}
is in the higher frequency range that is experimentally observed,
and then the increasing trends of ${\cal J}^{(1)}$ and ${\cal J}^{(2)}$
can be obtained by the NP approach
without recourse to the smooth interpolation,
which is necessary for the RPA.

\section{Summary}
\label{sec:sum}

In the this contribution I explained how the effects
of the pairing fluctuations appear in rapidly rotating nuclei.
By making use of the response function technique,
the correlation energy induced by the pairing fluctuations
can be evaluated within the RPA method.
The calculated RPA correlation energy is rather constant
as long as the static pairing gap is sizable,
but its absolute value decreases after the static gap is quenched.
In this way, the pairing fluctuations result in
dealignments of about a few units
with respect to the mean field calculation,
which makes the agreements with experimental data much better
in both normal deformed and superdeformed nuclei.
Thus, the effects of the pairing fluctuations are important
especially after the normal phase being realized at high-spin states.

\section*{Acknowledgements}
I express sincere gratitude to collaborators in the related works.
Especially, I am deeply in debt to Jerry D.~Garrett
for inspiring discussions and understanding experimental data,
who passed away in August, 1999.

%%%%%%%%%%%%%%%%%%%%%%%%%%%%%%%%%%%%%%%%%%%%%%%%%%%%%%%%%%%%%%5
% bibliography
%%%%%%%%%%%%%%%%%%%%%%%%%%%%%%%%%%%%%%%%%%%%%%%%%%%%%%%%%%%%%%5

\end{document}